\documentclass[aps,column,12pt,showpacs]{revtex4}
\usepackage{amsmath}
\usepackage{epsf}
\usepackage{epsfig}
\begin{document}
\title{NONEXTENSIVE STATISTICAL MECHANICS APPLICATION TO VIBRATIONAL
DYNAMICS OF PROTEIN FOLDING }
\author{Ethem Akt\"urk}
\email[E-mail: ]{eakturk@hacettepe.edu.tr}\affiliation{Department
of Physics Engineering, Hacettepe University, 06800 Ankara,
Turkey}
\author{Handan Ark{\i}n}
\email[E-mail: ]{handan@hacettepe.edu.tr}\affiliation{Department
of Physics Engineering, Hacettepe University, 06800 Ankara,
Turkey}
\begin{abstract}
The vibrational dynamics of protein folding  is analyzed in the
framework of Tsallis thermostatistics. The generalized partition
functions, internal energies, free energies and temperature factor
(or Debye-Waller factor) are calculated. It has also been observed
that the temperature factor is dependent on the non-extensive
parameter q which behaves like a scale parameter in the harmonic
oscillator model. As $q\rightarrow 1$, we also show that these
approximations agree with the result of Gaussian network model.

\end{abstract}

\maketitle


\section{Introduction}

The statistical mechanics paradigm based on Boltzmann-Gibbs
entropy which is great success looks after be incapable to deal
with many interesting physical systems\cite{NORMA}.  In  1988,
Tsallis advanced   a nonextensive   generalization  of
Boltzamann-Gibbs entropic measure. There are many applications for
Tsallis formalism such as the specific heat of the harmonic
oscillator~\cite{Ito}, one-dimensional Ising model~\cite{Andrade},
the   Boltzmann H-theorem~\cite{Mariz}, the Ehrenfest
theorem~\cite{aplastino}, quantum statistics~\cite{buyukkilic1},
paramagnetic systems~\cite{buyukkilic2},  Cercle
maps~\cite{tirnakli1},  Henon map~\cite{tirnakli2}, Haldane
exclusion statistics~\cite{Rajagapol}, q-expectation
value\cite{abe}, quantum mechanical treatment of
constraints\cite{bagci}. In this paper, we will focus on
thermodynamics and vibration properties of proteins in the
framework of a generalized thermostatistics. Before proceeding
with generalized gaussian network model, we described fundamental
definitions of Tsallis thermostatistics in Section II. We then
study generalized gaussian network model and result in section
III. Results will be discussed in section IV.

\section{Fundamental Definitions of Tsallis Thermostatistics}

The generalization    which   has been   successfully nonextensive
statistical   mechanics  is based  on  the  following expression
\cite{ts88}

\begin{eqnarray}
S_{q}=k\frac{1-Tr(\rho^{q})}{q-1}.
\end{eqnarray}
\begin{flushleft}
In the Tsallis statistics, the nonextensive canonical distribution
is given by
\end{flushleft}
\begin{equation}
\rho _{q}=\frac{1}{Z_{q}}[1-(1-q)\beta H]^{1/(1-q)},
\end{equation}
where
\begin{eqnarray}
Z_{q}=\int \prod_{n=1}^{N}dp_{n}dx_{n}[1-(1-q)\beta H]^{1/(1-q)},
\end{eqnarray}
is the partition function with $q\in {\mbox {\bf R}}$ and $\beta $
$\equiv $ $1/T$ (with $k=1$). Eq.(2) is obtained maximizing the
Tsallis entropy using lagrange multiply methods\cite{ts88,ct91}.
The internal energy is described by
\begin{eqnarray}
U_{q}=\int \prod_{n=1}^{N}dp_{n}dx_{n}\rho ^{q}H
\end{eqnarray}
where
\begin{eqnarray}
\int \prod_{n=1}^{N}{\mbox d}p_{n}{\mbox d}x_{n}\;\rho^{q}=1.
\end{eqnarray}
In Free energy notation, Eq.(4) appears as
\begin{eqnarray}
U_q = F_q -T \partial F_q /\partial T,
\end{eqnarray}
where $F_{q}$ is free energy is described as
\begin{eqnarray}
F_{q}=(Z_{q}^{q-1}-1)/[(1-q)\beta].
\end{eqnarray}
The expectation of an observable A is
\begin{equation}
A_{q}=\langle A \rangle_q =\int
\prod_{n=1}^{N}dp_{n}dx_{n}\rho^{q}.
\end{equation}
\section{Generalized Gaussian Network Model and Results}
We focus our discussion on the classical Tsallis statistics by
considering protein folding and harmonic approximation. Consider
the following Hamiltonian for Gaussian Network Model(GNM) based on
harmonic approximation\cite{bahar}
\begin{equation}
H=\sum_{i=1}^{N}\frac{p_{i}^{2}}{2m}+\frac{\gamma}{2}\Delta
R_{i}^{T}L\Delta R_{i},
\end{equation}
where the first term is the kinetic energy of the system and
$\gamma$ is the strength of the springs. In this model, springs
are assumed as homogeneous. ${ R}_i$ and $\Delta  R_{i}$ is also
defined as the equilibrium position and the displacement with
respect to $ R_{i}$ of the $i$-th C$_{\alpha}$ atoms. The model is
eventually described by the contact matrix $L$ with entries:
$L_{ij}=1$ if the distance $|{\bf R}_i - {\bf R}_j|$ between two
C$_{\alpha}$'s, in the native conformation, is below the cutoff
$R_0$, while is $0$ otherwise. From Eq.(3), partition function for
the Gaussian Network model\cite{bahar} which has above hamiltonian
is given by
\begin{eqnarray}
Z_{q}&=&\int \prod_{i=1}^{N}{\mbox d}p_{i}{\mbox d}\Delta
R_{i}\left[
1-(1-q)\beta \right.\nonumber \\
& & \times \left.\left(
\sum_{i=1}^{N}\frac{p_{i}^{2}}{2m}+\frac{\gamma}{2}\Delta
R_{i}^{T}L\Delta R_{i}\right) \right] ^{1/(1-q)}.
\end{eqnarray}
we introduce the variables $L=V^{T}\lambda V $, $ x=V \Delta R$
and thus $\Delta R^{T}L\Delta R=x^{T}\lambda x$ if $\lambda$ is
eigenvalues of the $L$ and $V$ is eigenvectors of $L$. We can
rewrite partition function as:
\begin{eqnarray}
Z_{q}&=&\int \prod_{i=1}^{N}{\mbox d}p_{i}{\mbox d} x_{i}\left[
1-(1-q)\beta \right.\nonumber \\
& & \times \left.\left(
\sum_{i=1}^{N}\frac{p_{i}^{2}}{2m}+\frac{\gamma}{2}\lambda_{i}x_{i}^{2}\right)
\right] ^{1/(1-q)}.
\end{eqnarray}
To calculate this integral, we definition new variables:
$y_{i}=[(1-q)\gamma\lambda{i}\beta /2]^{1/2}\ x_{i}$ and
$y_{N+i}=[(1-q)\beta /(2m)]^{1/2}p_{i}$ \cite{Lenzi}, where
$i=1,2,3,...,N$. Substituting these variables into $Z_{q}$ gives
\begin{eqnarray}
Z_{q}&=&\left \{ \prod_{i}^{N}\left[
\frac{2}{(1-q)\lambda_{i}\beta} (\frac {m}{\gamma})^{1/2}
\right] \right \} \nonumber \\
& &\times\int \prod_{n=1}^{2N}{\ \mbox d}y_{n}\left( 1-
\sum_{k=1}^{2N}y_{k}^{2}\right) ^{1/(1-q)}.
\end{eqnarray}
By using hyperspherical coordinates with $
u=(\sum_{n=1}^{2N}y_{n}^{2})^{1/2}$ and calculating the integral
over the angular variables\cite{Lenzi}, we obtain
\begin{eqnarray}
Z_{q}&=&\left \{ \prod_{n=1}^{N}\left[ \frac{2}{(1-q)\lambda_{n}\beta} (\frac {m}{\gamma})^{1/2}\right] \right \}\frac{\Omega _{2N}}{2}\nonumber \\
& &\times \int_{0}^{1}{\mbox d}u\;u^{N-1}\left( 1-u\right)
^{1/(1-q)}.
\end{eqnarray}
By using definition of the solid angle\cite{Greiner} $\Omega
_{2N}=2\pi ^{N}/\Gamma (N)$ and  using integral representation of
Euler beta function\cite{Arfken}, we obtained that
\begin{eqnarray}
Z_{q}&=&\left \{ \prod_{n=1}^{N}\left[
\frac{2\pi}{(1-q)\lambda_{n}\beta} (\frac
{m}{\gamma})^{1/2}\right] \right \}\frac{\Gamma \left(
\frac{1}{1-q}+1\right)
}{\Gamma \left( \frac{1}{1-q}+1+N\right) }\nonumber \\
&=& \left[\left(\frac{2-q}{1-q}\right)_N\right]^{-1}
\prod_{n=1}^{N}\left[ \frac{2\pi}{(1-q)\lambda_{n}\beta} (\frac
{m}{\gamma})^{1/2}\right],
\end{eqnarray}
where $(a)_n= a(a+1)(a+2)...(a+n-1)=\frac{(a+n-1)!}{(a-1)!}$ is
the Pochhammer symbol \cite{Arfken}. Free energy of the system is
given by
\begin{eqnarray}
F_{q}=-\frac{1}{\beta}\ln_{q}Z_{q},
\end{eqnarray}
and
\begin{eqnarray}
F_{q}=-\frac{1}{\beta}\ln_{q}\left[\left(\frac{2-q}{1-q}\right)_N\right]^{-1}
\prod_{n=1}^{N}\left[ \frac{2\pi}{(1-q)\lambda_{n}\beta} (\frac
{m}{\gamma})^{1/2}\right],
\end{eqnarray}
and then
\begin{eqnarray}
F_{q}=-\frac{1}{\beta}\ln_{q}\left[\left(\frac{2-q}{1-q}\right)_N\right]^{-1}
\left[ \frac{2\pi}{(1-q)\beta } (\frac
{m}{\gamma})^{1/2}\right]^{N} (\det{(L^{-1})})^{1/2}.
\end{eqnarray}
From Eq.(6), the generalized internal energy can be calculated as
\begin{eqnarray}
U_q&=&\frac{N}{\beta}\; Z_{q}^{1-q}\frac{N}{\beta}\left \{ \left [
\frac{2\pi }{(1-q) \beta}(\frac {m}{\gamma})^{1/2}\right]
^{N}(\det{(L^{-1})})^{1/2}\left[\left(\frac{2-q}{1-q}\right)_N\right]^{-1}
\right \}^{1-q}.
\end{eqnarray}
Hence, we obtained thermodynamic properties of the generalized
Gaussian network model. On the other hand, the most properties of
Gaussian network model is called temperature factor. The
crystallographic temperature factors or B-factors, which is
defined as intrinsic fluctuations of the atoms in crystal, is
directly related to this fluctuations. The X-ray crystallographic
temperature factors are used by the mean square fluctuation of
C$_\alpha$ atoms around their native positions\cite{bahar}. The
generalized temperature factor is expressed as
\begin{eqnarray}
B_{iq}(T) = \frac{8\pi^2}{3}\langle x_{i}x_{i}\rangle_{q},
\end{eqnarray}
with $\langle \cdot \rangle$ indicating the thermal average. Where
$\langle x_{i}x_{j}\rangle_{q}$ is defined by
\begin{eqnarray}
\langle x_{i}x_{j}\rangle_{q}=\frac{1}{Z_{q}^{q}}\int Dx Dp
 x_{i}x_{j}[1-(1-q)\beta H]^\frac{q}{1-q},
\end{eqnarray}
or
\begin{eqnarray}
\langle x_{i}x_{j}\rangle_{q}=\frac{1}{Z_{q}^{q}}\int Dx Dp
 x_{i}x_{j}\left[1-(1-q)\beta
\sum_{i=1}^{N}\frac{p_{i}^{2}}{2m}+\frac{\gamma}{2}\lambda_{i}x_{i}^{2}
\right]^\frac{q}{1-q}.
\end{eqnarray}

 In the GGNM (generalized gaussian network model)  approximation, this
average is easily carried out, because amounts to a nonGaussian
integration, and B-factors can be expressed in terms of the
diagonal part of the inverse of the matrix

\begin{eqnarray}
\langle x_{i} \cdot x_{j} \rangle_{q} = \frac{3 k_B T}{\gamma}
[\frac{L^{-1}_{ij}}{1+(1-q)(N+1)}]
\end{eqnarray}.

\begin{figure}[htbp]
\epsfig{file=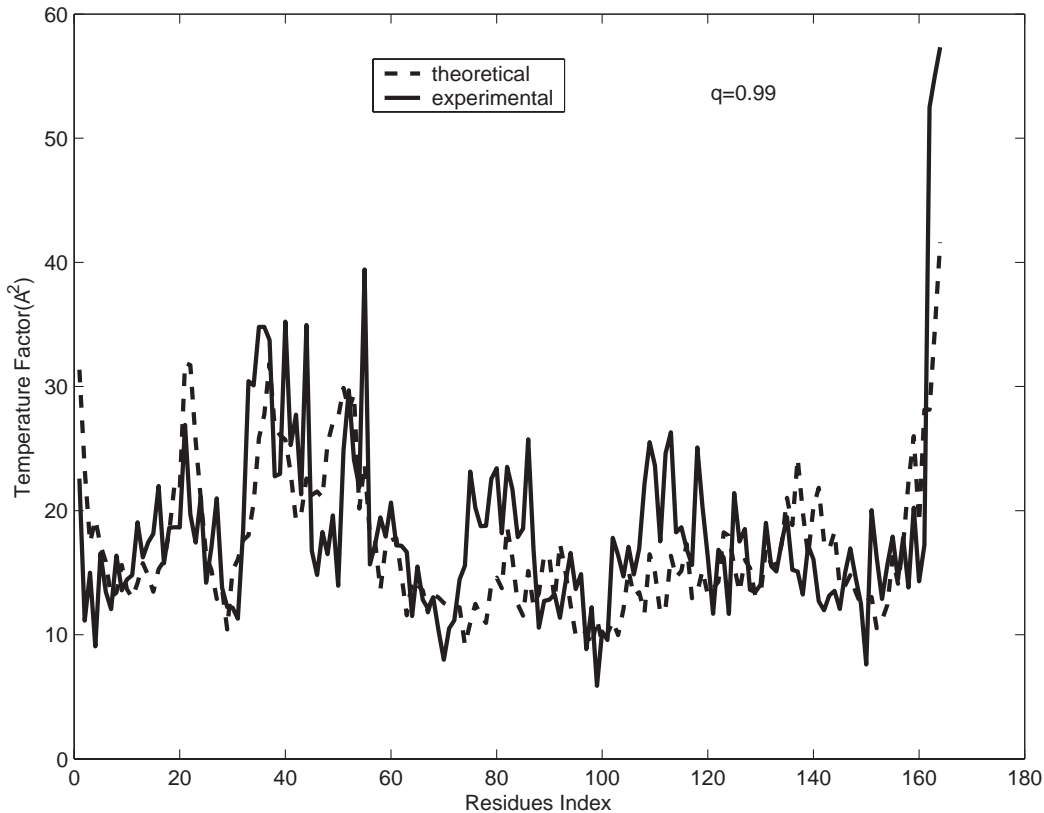,width=14cm,height=11cm}\caption{Temperature
factors for T4 lysozyme}\label{eps1}
\end{figure}

\begin{figure}[htbp]
\epsfig{file=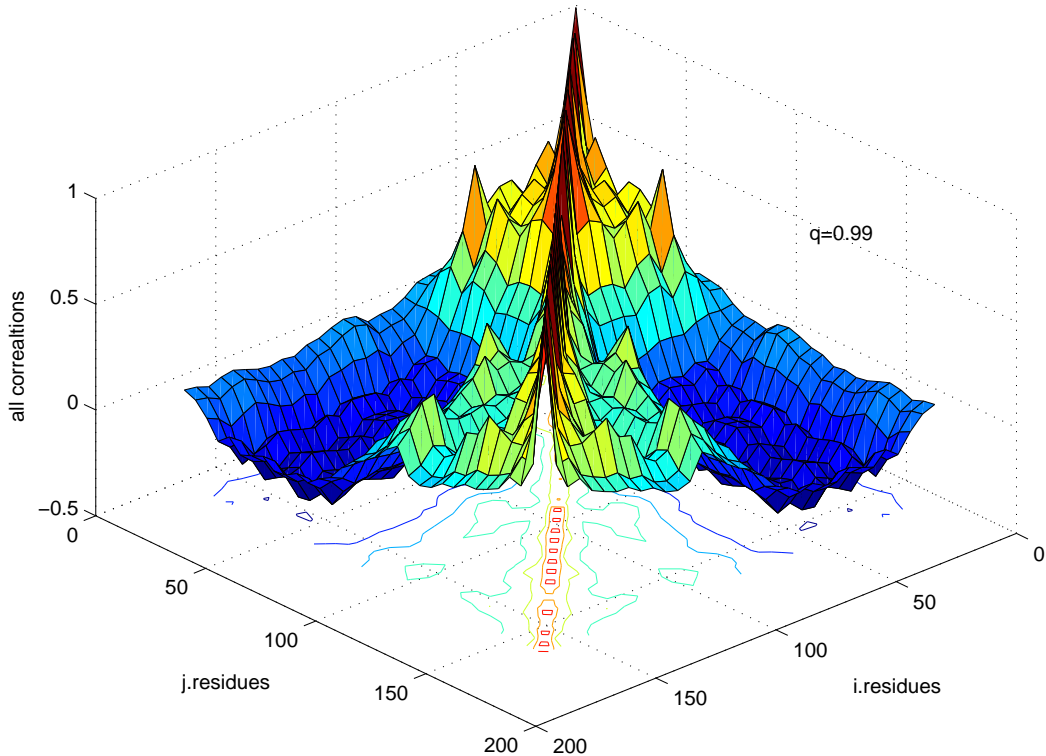,width=14cm,height=11cm}\caption{Cross correlation between pair residues 
for T4 lysozyme}\label{eps2}
\end{figure}

Substituting Eq. 20 into Eq.17 gives
\begin{eqnarray}
B_{iq}(T) =\frac{8\pi^2 k_B T}{\gamma}
[\frac{L^{-1}_{ii}}{1+(1-q)(N+1)}].
\end{eqnarray}
If $q\rightarrow 1$, it is calculated that same result of the
gaussian network model as\cite{wagner}
 \begin{eqnarray} B_i(T) =\frac{8\pi^2 k_B T}{\gamma} [L^{-1}_{ii}]
\end{eqnarray}
For the numerical results, we choice a reasonable cutoff distance
including all residues pairs within a first interaction shell is
7.0 \AA. We display in Fig.\ref{eps1} temperature factors for T4
lysozyme (or 3LZM pdb) as a function of residue index at q=0.99. 
Bold and dashed curves depict the experimental and theoretical
results, respectively. The theoretical curve is normalized by
taking $\gamma(\AA^{-2})=2.06 k_{B}T$, therefore as to match the
area enclosed by the two curves. The agreement between the
theoretical and experimental curves is remarkable. In 
Fig.\ref{eps2}, we also display to behavior cross correlation between 
i.residues and j.residues of all modes at q=0.99.   
\section{Conclusion}
In this study, we obtained  temperature factor or Beta factor
using by Tsallis statistical mechanics. In this framework, by the
approximate scheme which we have used here, closed and analytical
expressions for the generalized gaussian network model  have been
derived for harmonic oscillator. In addition to this, the
temperature factor is q dependent. Since the temperature factor
must be positive, q must be less than unity. This also shows that
q must be less than unity for harmonic oscillator system. As can
be seen in Fig.\ref{eps1}, at q=0.99, the beta factors which are
obtained from experimental and numerical calculation are in
agreement with each other. The more the q parameter decreases, the
more the temperature factor disagrees with experimental data.
Therefore, for q values in the lower vicinity of 1, the
experimental data are in agreement with numerical calculations. In
this system, as $q\rightarrow 1$, it is observed that, both of the
above expressions reduce to the results obtained in the standard
Boltzmann-Gibbs statistics.

\section{Acknowledgements}
H.A.\ acknowledges support by the T\"{U}B\.ITAK under the project
number 104T150 and from  T\"{U}BA under the Programme to Reward
Successful Young Scientists.


\end{document}